# Direct generation of eye-safe single- and dual-vortex lasers via off-axis pumping of the active medium

Xiaoxu Huang, Shengwei Cui, Xiaofeng Guan, Bin Xu[a)], Huiying Xu, and Zhiping Cai

*Department of Electronic Engineering, Xiamen University, Xiamen 361005, China*

A simple and high-efficiency method for direct generation of all-solid-state single- and dual-vortex lasers operating at eye-safe wavelengths is reported in a compact diode-end-pumped Nd:YAG laser cavity. By off-axis pumping of the Nd:YAG in two orthogonal directions, the original cavity mode, fundamental Gaussian mode ($TEM_{00}$), is directly transformed into first-order Laguerre-Gaussian ($LG_{01}$) mode with single- or dual-vortex structures depending on the degree of the off-axis pumping. Moreover, the single- and dual-vortex lasers can be produced in simultaneous eye-safe dual-wavelength operation with the aid of an intracavity etalon. At the same time, a novel method to determine the handedness of achieved vortex lasers by using a plane-concave mirror with suitable coating is also proposed. This method can be perfectly applied to generate vortex lasers at other emissions bands for various applications by using different laser gain mediums. This work paves a general way to simple and efficient realization and determination of vortex lasers.

The singularity of the radiation field is identified by the term "optical vortex", which is characterized by helical phase fronts. Such optical vortex beam with ring-shaped intensity profile has become more and more attractive because it is capable of providing orbital angular momentum (OAM) of $\pm l\hbar$ ($l$: integer) for each photon corresponding to azimuthal dependence of the phase front, $\exp(\pm il\phi)$. Based on this distinct OAM property, optical vortex beams could have a variety of applications, such as in particle manipulation [1-3], quantum information [4], material processing [5,6], optical microscopy [7] and large capacity of optical communication systems [8,9].

To generate optical vortex beam with phase singularities artificially from an initial plane or Gaussian wave, a number of experimental techniques have been elaborated, which in general involve the utilizations of some particular optical elements, such as spiral phase plate [10], astigmatic lenses [11], optical wedges [12], cylindrical lens pair [13], spatial light modulators [14] and computer-generated holographic converters [15]. However, first of all, the introductions of these optical devices make the systems complicated, difficult in alignment and high cost. Moreover, some of these methods suffer from intrinsic drawbacks originating from the low damage threshold of these devices, e.g. spatial light modulators. Recently, vortex lasers generation based on intracavity method, i.e. the so-called direct generation, have drawn rising attention. For example, in 2012, $LG_{01}$ laser modes were directly generated in a diode-pumped Nd:LYSO laser with doughnut-shaped beam of the pump source under the help of thermal lensing effect of the laser gain medium [16]. In 2013, using a 808-nm pump beam reconstructed by a capillary (hollow-core) fiber into a doughnut-shaped beam, Kim et al. [17] reported an actively Q-switched Nd:YAG

---

[a)] Author to whom correspondence should be addressed.  Electronic mail: xubin@xmu.edu.cn.

vortex laser with the aid of an etalon and a Brewster plate to ensure a particular helical wavefronts. In 2014, utilizing the combination of doughnut-shaped pump beam and thermal lensing effect of the laser gain medium, graphene-based passively Q-switched Nd:LYSO vortex laser was also demonstrated [18]. In fact, recently, intracavity vortex laser generation by using doughnut-shaped pump beam have become common and popular in such researches [19,20]. The reason is clear and fundamental: To match doughnut-shaped pump beam, the laser resonator is prone to generate ring-shaped laser mode. However, on the one hand, according to these previous publications, it does not lead to vortex laser automatically for pumping with doughnut-shaped beam, thermal lensing effect must occur to transform cavity mode into LG mode (see e.g. Ref. [16]), which indicates that the LG mode cannot be operated at threshold or even at low output power. Moreover, laser materials with weak thermal lensing effect will be ineffective on LG mode generation. On the other hand, in fact, pump source with doughnut-shaped beam is not standard and reshaping the pump beam into the required intensity profile not only complicates the system but also reduces the pump conversion efficiency. Intensive research on the development of simple and reliable methods for direct vortex laser generation is still on demand.

For solid-state lasers, HG (Hermite-Gaussian) modes can be generated easily by off-axis pumping of laser gain medium, i.e. parallel displacing the pump beam from the optical axis of the laser resonator [13,21]. In this work, we realized direct generation of vortex lasers via off-axis pumping of the gain medium by rotating the gain medium (not parallel displacing the pump beam) in two orthogonal directions. Basically, this method does not introduce other optics inside or outside the laser cavity and this method has also been experimentally confirmed to be very efficient. Last but not least, a novel method by using a simple plane-concave mirror instead of the commonly used Mach-Zehnder interferometer to check the handedness of helical wavefronts has also been presented in this work.

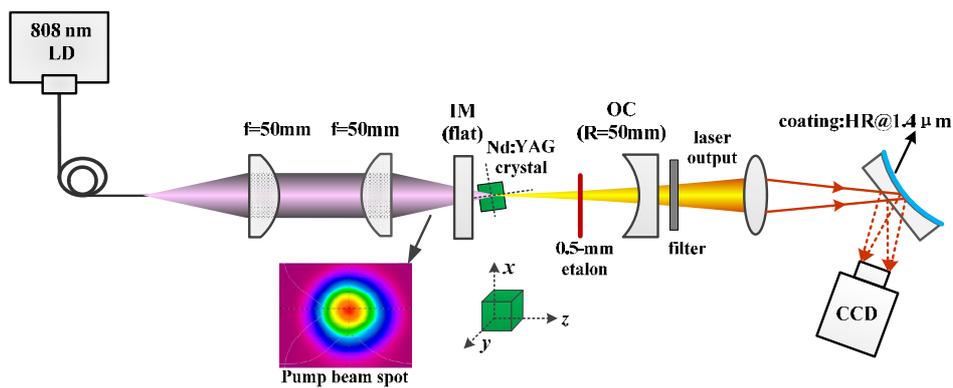

FIG. 1. Schematic of diode-pumped Nd:YAG vortex laser.

To realize direct vortex laser generation, a simple and compact diode-pumped Nd:YAG laser with two-mirror configuration was arranged, as shown in Fig. 1. The pump source was a fiber-coupled diode laser emitting at 808 nm with core diameter of 400 μm and numerical aperture of 0.22. The noticeable difference of the present vortex laser experiment lies in the utilization of a commoner and standard pump source with $TEM_{00}$-mode intensity profile, instead of with ring-shaped pump beam. The pump beam was collimated and focused by two doublet lenses both with focal lengths of 50 mm. The used laser gain medium was a Nd:YAG crystal with dopant concentration of 0.5at%, cross section of 3×3 $mm^2$ and thickness of 7 mm. The end-face mirrors of the laser resonator consisted of a flat input mirror (IM) and a 50-mm (curvature radius) output coupler (OC). The IM has a high transmission of about 87% at pumping wavelength and a high reflection of about 99.8% at 1.4 μm. The IM also has high transmissions of more than 90% at 1.06 and 1.1 μm, as well as transmission of more than 40% at 1.3 μm to suppress these high-gain emission lines. The OC has transmissions of about 1.79% and 2.39% at 1412 and 1442 nm.

At first, the laser cavity was optimized to realize a high-power laser output with cavity length of about 44 mm. Fig. 2 shows the dependence of laser output power on the absorbed power of the laser crystal. A maximum output power up to 1.55 W was achieved at 1442 nm with threshold of 1.41 W in absorbed power and slope efficiency of about 20.5%. The present laser results are comparable to that reported in Ref. [22] with similar laser output power and efficiency when considering the same absorbed power, while our laser threshold is lower than the reported results. Beam spot of the output laser is also shown as an inset in Fig. 2, which clearly shows a Gaussian $TEM_{00}$ mode.

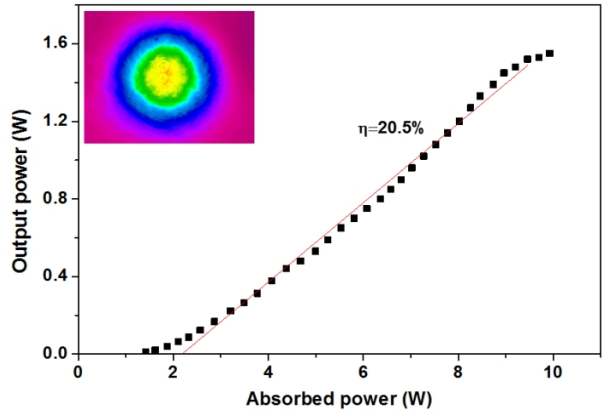

FIG. 2. Output power characteristic of Nd:YAG laser at 1442 nm with inset showing $TEM_{00}$ mode.

In the following, the laser gain medium was rotated slightly in *xz* plane to force off-axis pumping. By doing this step, a $HG_{01}$/$HG_{10}$ mode in diagonal direction having petal-like pattern with two lobes was achieved, as Fig. 3(a) shown. The field distribution of the π/4 rotated $HG_{01}$/$HG_{10}$ mode relative to the *x*- or *y*-axis is equivalent to the sum of two $HG_{01}$ and $HG_{10}$ modes. The importance for realizing mode conversion from HG to LG lies in a π/2 Gouy phase [11], which can be realized by passing the beam through a mode converter to make it astigmatic, e.g. a pair of cylindrical lens [11,13] or equivalent operation to form such astigmatism. Further slightly rotating the Nd:YAG crystal in *yz* plane was conducted and consequently the $HG_{01}/HG_{10}$ mode was observed to transform into a transverse mode with donut-shaped intensity profile (see Fig. 3(b)). Since such doughnut beam could be coherent or incoherent superposition of two petal beams [19,23], it is necessary to confirm the vortex property of the achieved doughnut beam. Doughnut beam formed by incoherent superposition leads to an unclear spiral pattern in the interference patterns, however clear and stable helical wavefronts with well-determined handedness can be observed for coherent superposition of two petal beams, i.e. pure LG mode [19].

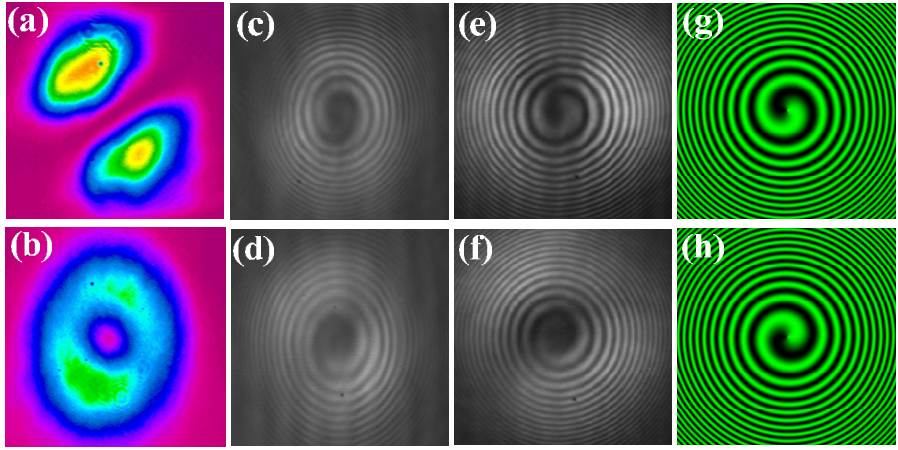

FIG. 3. (a) $HG_{01}/HG_{10}$ mode, (b) doughnut beam, (c) clockwise interference fringe, (d) anti-clockwise interference fringe, (e) enhanced clockwise interference fringe, (f) enhanced anti-clockwise interference fringe, (g) simulation of the clockwise interference fringe and (h) simulation of the anti-clockwise interference fringe.

Mach-Zehnder interferometer is still relatively complicated system in spite of that it has been often used to check the handedness of helical wavefronts. Recently, Khajavi et al. [24] reported a new method to determining topological charge by using a commercial lateral shearing interferometer based on a wedged optical flat. In this work, we presented a simpler and cost-effective method to realize the same goal in the absence of a Mach-Zehnder interferometer, as shown in Fig. 1. The output laser was injected into a plane-concave mirror after focused by a positive lens. The plane face of the mirror was uncoated and therefore only Fresnel reflection should be considered. The concave of the mirror was high-reflection coated at 1.4 μm. Thus, the optical fields of plane reflection and spherical surface reflection interfered at where a CCD was placed. Fig. 3(c) shows the interference pattern of the two optical fields, which is a spiral pattern in anti-clockwise direction. The handedness of the helical wavefronts can be shifted by slightly tilting the Nd:YAG a little more, as Fig. 3(d) shown that the spiral pattern in clockwise direction can also be observed. The interference patterns shown in Fig. 3(c) and 3(d) were pretty stable and the handedness can be determined quite well, and therefore the present uniform donut-shaped beam can be confirmed to be $LG_{01}$ mode with high purity. The qualities of the interference patterns can be further improved by making the vortex laser system more stable with less longitudinal modes. Therefore, a 0.5-mm-thick glass plate acting as F-P etalon was inserted into the laser resonator, as shown in Fig. 1. After finely tilting the etalon with a small and suitable angle, enhanced interference patterns with higher contrasts were achieved, as shown in Fig. 3(e) and 3(f), which indicated that the purity of the $LG_{01}$ mode was further improved. On the one hand, it is due to the etalon that reduced the existing longitudinal modes inside the laser resonator. On the other hand, optical components in the laser resonator are positioned symmetrically to the propagating direction, providing indistinguishable resonating loss to two $LG_{01}$ modes with opposite helicity, which degrades the purity and therefore contrast of the $LG_{01}$ mode. The introduction of etalon at specific angle may break the spiral propagation symmetry of two LG mode beams with opposite helical wavefronts, resulting in the selection of the wavefront handedness and further leading to enhanced contrast of the interference patterns [19].

The field amplitude of LG mode can be written by [25]

$$E_{LG}(r,\phi,z) \propto \exp(-ikr^2/2R)\exp(-r^2/w^2) \times \exp[-i(2p+l+1)\psi]\exp(-il\phi) \\ \times (-1)^p (r\sqrt{2}/w) \times L_p^l(2r^2/w^2) \quad (1)$$

where $\omega(z)$ is spot size, $R(z)$ is the wavefront curvature, $\psi(z)$ is the Gouy phase of LG mode, z is the distance from the beam waist, $r$ is the radius, $\phi$ is the azimuthal angle, $p$ and $l$ are the radial and azimuthal indexes of the mode respectively and $L_p^l(x)$ is the generalized Laguerre polynomial. We can simulate the interference patterns by using the phase term of expression (1), as shown in Fig. 3(g) and 3(h).

The generation of $LG_{01}$ mode can be qualitatively analyzed by threshold condition. As we know, LG mode, like HG mode, is the eigenmode of the laser resonator. For end-pumped solid-state lasers, according to the modeling of Fan and Byer [26], thresholds, $P_{th}$, with respect to different transverse modes can be modified for four-level lasers,

$$P_{th} = \frac{h\nu_p \delta}{2\sigma l t_f \eta_p \eta_a} \frac{1}{\int_{cavity} r_0(r,z) s_0(r,z) dV} \quad (1)$$

where $\delta$ is the round-trip loss of laser resonator, $\sigma$ is the emission cross section of the Nd:YAG crystal, $l$ is the length of the Nd:YAG, $t_f$ is the lifetime of the upper laser manifold, $\eta_p$ is the pump quantum efficiency and $\eta_a$ is the absorption ratio of the pump power. The $r_0(r,z)$ and $s_0(r,z)$ are the normalized intensity distributions of pump beam and cavity modes. In general, $TEM_{00}$ mode oscillates firstly because of its lowest threshold. By rotating the laser gain medium in *xz* plane firstly to enforce an off-axis operation, $TEM_{00}$ mode was suppressed because of an increased threshold for it, which originated from resonator mode mismatching and therefore increased intracavity loss for $TEM_{00}$ mode. Thus, as a consequence, $HG_{01}/HG_{10}$ mode exhibiting lower threshold oscillated before $TEM_{00}$ mode. Similarly, $LG_{01}$ mode started to oscillate subsequently since the laser resonator enforced additional loss to $HG_{01}/HG_{10}$ mode by further rotating the Nd:YAG crystal in *yz* plane.

Fig. 4(a) shows that the maximum output power of the single vortex $LG_{01}$ laser reached 1.48 W, i.e. a power attenuation of less than 5% compared to the above $TEM_{00}$ mode. Moreover, threshold of the $LG_{01}$ laser was close to the $TEM_{00}$ laser. During all the measurements of the output powers, the output laser was monitored to be vortical. That is to say, the increasing thermal lensing effect inside the laser crystal has not influenced the vortex output, which made this method for vortex generation more reliable and of advantage. Linear fitting these data, the slope efficiency of the single vortex laser was found to be about 18.6%. In addition, under this situation, the lasing wavelength was also found to be about 1442 nm, as shown in Fig. 4(b).

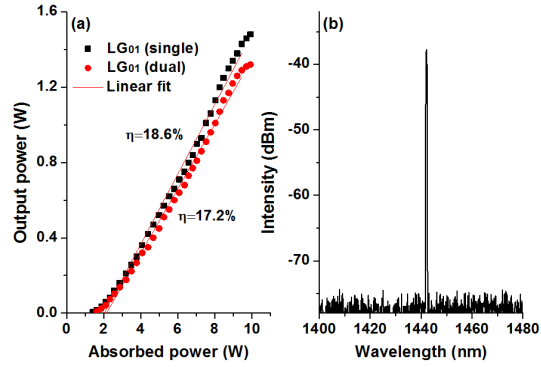

FIG. 4. (a) Output power characteristics of Nd:YAG laser with single $LG_{01}$ mode and dual $LG_{01}$ mode; (b) the corresponding output spectrum showing peak wavelength at ~1442 nm.

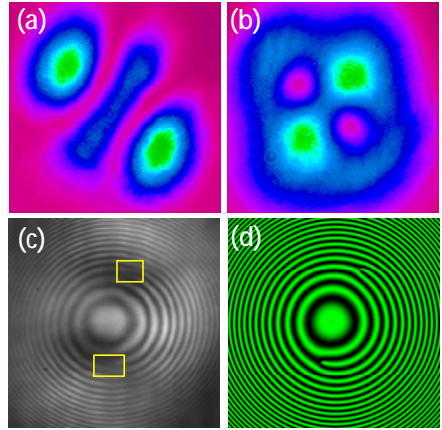

FIG. 5. (a) Gaussian beam spot of $HG_{02}/HG_{20}$ mode, (b) vortex laser with dual $LG_{01}$ modes, (c) interference fringe of the dual vortex lasers and (d) simulation on the interference fringe of the dual vortex lasers.

In this work, simultaneous dual vortex laser generation can also be realized. As Fig. 5(a) shown, similar to the above descriptions, at first, $HG_{02}/HG_{20}$ mode was achieved by rotating the Nd:YAG in *xz* plane to a larger angle. After, tilting the Nd:YAG in *yz* plane again, we then attained the laser output with two vortices, still in $LG_{01}$ mode for each one (see Fig. 5(b)). Fig. 5(c) shows the interference pattern of the dual vortices with spherical surface wave, which contains two forks in symmetrical positions to the central fringe. Note that the interference pattern was obtained after the insertion of the 0.5-mm etalon again. Simulation of the interference pattern is shown in Fig. 5(d) by adding a new vortex phase term in expression (1), which shows good agreement with the actual interference pattern. The output powers of the dual-vortex laser were also plotted in Fig. 4(a), which shows the maximum output power and slope efficiency reduced to 1.32 W and 17.2%, respectively. Note that the twin-vortex laser operated still at about 1442 nm.

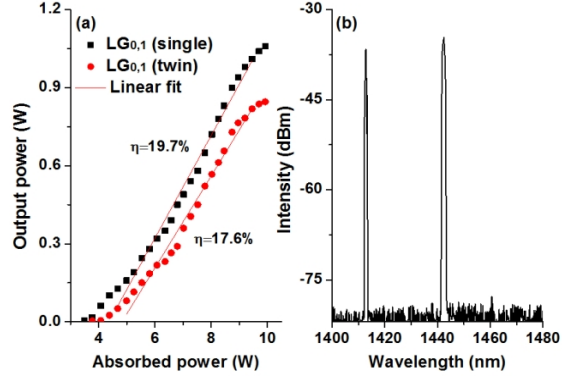

FIG. 6. (a) Output power characteristics of dual-wavelength Nd:YAG laser with single and dual $LG_{0,1}$ mode; (b) the corresponding output spectrum showing a dual-wavelength laser at ~1412 and ~1442 nm.

At last, simultaneous dual-wavelength and dual-vortex laser was also realized by inserting a 0.13-mm-thick etalon into the laser resonator to modulate the intracavity loss with respect to different emission lines. As Fig. 6(a) shown, for single vortex case, the maximum output power reduced to 1.06 W and the threshold increased to 3.49 W, thus leading to a slope efficiency of about 19.7%. For dual-vortex case, a maximum output power of 0.85 W, a threshold of 3.78 W and a slope efficiency of 17.6% were obtained. Both of the single and dual vortex lasers operated in simultaneous dual-wavelength mode at 1412 and 1442 nm because of the etalon, as Fig. 6(b) shown.

In conclusion, 1.55-W $TEM_{00}$-mode eye-safe Nd:YAG laser at 1442 nm with slope efficiency of 20.5% was achieved using a simple and compact diode-end-pumped two-mirror laser cavity. Based on the $TEM_{00}$ mode laser, by rotating the Nd:YAG crystal, $LG_{01}$ laser with single vortex was then obtained with maximum output power of 1.48 W and slope efficiency of 18.6%. Further tilting the Nd:YAG crystal, laser with dual-vortex structure can also be generated with maximum output power of 1.32 W and slope efficiency of about 17.2%. The present vortex laser can also be operated in dual-wavelength regime at 1442 nm and 1412 nm by inserting an etalon inside the laser cavity. Under this situation, i.e. dual-wavelength lasing, the maximum output powers for single- and dual-vortex lasers reached 1.06 W and 0.85 W, respectively. What's more, in this work, we have presented a new and simple method to determine the handedness of the vortex by only using a plane-concave mirror with suitable coating. Both methods show very good generalities, which can be safely applied to other laser systems for vortex laser generation and determination. As a proof, using the same methods, we have also successfully obtained single- and dual-vortex Nd:YVO$_4$ lasers at 1.3 μm.

[1]N. B. Simpson, L. Allen, M. J. Padgett, "Optical tweezers and optical spanners with Laguerre-Gaussian modes," J. Mode. Opt. 43 (12), 2485-2491 (1996)

[2]M. Padgett and R. Bowman, "Tweezers with a twist," Nat. Phot. 5, 343-348 (2011)

[3]R. P. Lopez, U. Ruiz, V. Arrizon, and R. R. Garica, "Optical manipulation using optimal annular vortices," Opt. Lett. 41 (17), 4138-4141 (2016)

[4]A. Mair, A. Vaziri, G. Weihs, and A. Zeilinger, "Entanglement of the orbital angular momentum states of photons," Nature 412, 313-316 (2001)

[5]K. Sueda, G. Miyaji, N. Miyanaga and M. Nakatsuka, "Laguerre-Gaussian beam generated with a multilevel spiral phase plate for high intensity laser pulses," Opt. Express 12 (15), 3548-3553 (2004)

[6]C. Hnatovsky, V. G. Shvedov, W. Krolikowski, and A. V. Rode, "Materials processing with a tightly focused femtosecond laser vortex pulse," Opt. Lett. 35 (20), 3417-3419 (2010)

[7]X. P. Li, P. Venugopalan, H. R. Ren, M. H. Hong, and M. Gu, "Super-resolved pure-transverse focal fields with an enhanced energy density through focus of an azimuthally polarized first-order vortex beam," Opt. Lett. 39 (20), 5961-5964 (2014)

[8]J. Wang, J. Y. Yang, I. M. Fazal, N. Ahmed, Y. Yan, H. Huang, Y. X. Ren, Y. Yue, S. Dolinar, M. Tur, and A. E. Willner, "Terabit free-space data transmission employing orbital angular momentum multiplexing," Nat. Phot. 6, 488-496 (2012)

[9]I. M. Fazal, N. Ahmed, J. Wang, J. Y. Yang, Y. Yan, B. Shamee, H. Huang, Y. Yue, S. Dolinar, M. Tur, and A. E. Willner, "2Tbit/s free-space data transmission on two orthogonal orbital-angular-momentum beams each carrying 25 WDM channels," Opt. Lett. 37 (22), 4753-4755 (2012)

[10]V. V. Kotlyar, A. A. Almazov, S. N. Khonina, and V. A. Soifer, "Generation of phase singularity through diffracting a plane or Gaussian beam by a spiral phase plate," J. Opt. Soc. Am. A 22 (5), 849 (2005)

[11]M. W. Beijersbergen, L. Allen, H. E. L. O. van der Veen, and J. P. Woerdman, "Astigmatic laser mode converters and transfer of orbital angular momentum," Opt. Commun. 96, 123-132 (1993)

[12]Y. Izdebskaya, V. Shvedov, and A. Volyar, "Generation of higher-order optical vortices by a dielectric wedge," Opt. Lett., 30, 2472–2474 (2005)

[13]C. Y. Lee, C. C. Chang, C. Y. Cho, P. H. Tuan, and Y. F. Chen, "Generation of Higher Order Vortex Beams From a $YVO_4$/Nd:$YVO_4$ Self-Raman Laser via Off-Axis Pumping With Mode Converter," IEEE J. Sel. Topics Quantum Electron. 21 (1), 1600305 (2015)

[14] N. Matsumoto, T. Ando, T. Inoue, Y. Ohtake, N. Fukuchi, and T. Hara, "Generation of high-quality higher-order Laguerre-Gaussian beams using liquid-crystal-on-silicon spatial light modulators," J. Opt. Soc. Am. A 25 (7), 1642-1651 (2008)

[15] N. R. Heckenberg, R. McDuff, C. P. Smith, and A. G. White, "Generation of optical phase singularities by computer-generated holograms," Opt. Lett. 17, 221-223 (1992)

[16] Y. G. Zhao, Z. P. Wang, H. H. Yu, S. D. Zhuang, H. J. Zhang, X. D. Xu, J. Xu, X. G. Xu, and J. Y. Wang, "Direct generation of optical vortex pulses," Appl. Phys. Lett. 101, 031113 (2012)

[17] D. J. Kim, J. W. Kim, and W. A. Clarkson, "Q-switched Nd:YAG optical vortex lasers," Opt. Express 21 (24), 29449 (2013)

[18] Y. T. Ding, M. M. Xu, Y. G. Zhao, H. H. Yu, H. J. Zhang, Z. P. Wang, and J. Y. Wang, "Thermally driven continuous-wave and pulsed optical vortex," Opt. Lett. 39 (8), 2366-2369 (2014)

[19] D. J. Kim and J. W. Kim, "Direct generation of an optical vortex beam in a single-frequency Nd:YVO$_4$ laser," Opt. Lett. 40 (3), 399-402 (2015)

[20] Y. G. Zhao, Q. Y. Liu, W. Zhou, and D. Y. Shen, "~1 mJ pulsed vortex laser at 1645 nm with well-defined helicity," Opt. Express 24 (14), 15596 (2016)

[21] Y. F. Chen, T. M. Huang, C. F. Kao, C. L. Wang, and S. C. Wang, "Generation of Hermite-Gaussian Modes in Fiber-Coupled Laser-Diode End-Pumped Lasers," IEEE. J. Elec. Quan. 33, 1025 (1997)

[22] H. N. Zhang, X. H. Chen, Q. P. Wang, X. Y. Zhang, J. Chang, L. Gao, H. B. Shen, Z. H. Cong, Z. J. Liu, X. T. Tao, and P. Li, "High efficiency Nd:YAG ceramic eye-safe laser operating at 1442.8 nm," Opt. Lett. 38 (16), 3075-3077 (2013)

[23] I. A. Litvin, S. Ngcobo, D. Naidoo, K. A. Ameur, and A. Forbes, "Doughnut laser beam as an incoherent superposition of two petal beams," Opt. Lett. 39 (3), 704-707 (2014)

[24] B. Khajavi, E. J. Galvez, "Determining topological charge of an optical beam using a wedged optical flat," Opt. Lett. 42 (8), 1516-1519 (2017)

[25] M. J. Padgett, and L. Allen, "The Poynting Vector in Laguerre-Gaussian Laser Modes," Opt. Commun. 121, 36-40 (1995)

[26] T. Y. Fan, R. L. Byer, "Modeling and CW operation of a quasi-three level 946nm Nd:YAG laser," IEEE J. Quan. Elec. 23 (5), 605-612 (1987)